\documentclass[aps,prb,showpacs,floatfix,twocolumn,superscriptaddress]{revtex4-1}
\usepackage{graphicx}
\usepackage{bm,amsmath}
\usepackage{dcolumn}
\begin{document}

\title{Hardcore bosons on the dual of the bowtie lattice}

\author{Wanzhou Zhang, Laixi Li, and Wenan Guo}
\affiliation{ Physics Department, Beijing Normal University, Beijing 100875, China}
\date{\today}
\begin{abstract}
We study the zero temperature phase diagram of hardcore bosons on the dual of the bowtie lattice. Two types of striped diagonal 
long-range order (striped order I and striped order II) are discussed. 
A state with type-II striped order and superfluidity is found, even without nearest-neighbor repulsion.
The emergence of such a state is due to the inhomogeneity and the anisotropy of the lattice structure. However, neither the translational 
symmetry nor the symmetry between sublattices of the original lattice is broken.  In this paper, we restrict  a 'solid state' of lattice
bosons as a diagonal long-range ordered state breaking either the translational symmetry of the original lattice or the symmetry of
different sublattices. We thus name such a phase a striped superfluid phase (SSF). 
In the presence of a nearest-neighbor repulsion, we find two striped charge density wave phases(SCDW I and II) with boson 
density $\rho=1/2$ (with striped order I)
and $\rho=2/3$ (with striped order II) respectively, when the hopping amplitude is small compared with the repulsion. 
The SCDW I state is a solid, in which the translational symmetry of the original lattice is broken. 
We observe a rather special first-order phase transition showing an interesting multi-loop hysteresis phenomenon 
between the two SCDW phases when the hopping term is small enough. This can be accounted for by the special degeneracy of the 
ground states near the classical limit.  The SSF re-appears outside the two SCDW phases. 
The transition between the SCDW I and SSF phases is first order, while the transition between SCDW II and SSF phases is 
continuous.  We find that the superfluid stiffness is anisotropic in the SSF states with and without repulsion.
In the SSF with repulsion, the superfluid stiffness is subject to different types of anisotropy in the region near half 
filling and above 2/3-filling.

\end{abstract}

\maketitle

The supersolid, a novel quantum state with diagonal simultaneous long-range order(DLRO) and off-diagonal long-range order(ODLRO), 
was introduced nearly half a century ago\cite{onsa}.
This exotic quantum state has attracted considerable  research interests in recent years.
Owing to the fast development of laser cooling techniques, various optical lattices can be realized\cite{optlatt}, which
makes the investigation of supersolidity of bosons on discrete lattices more realistic. 
For softcore bosons, the supersolid phase emerges by the "defect condensation" mechanism, where
doped bosons(holes) act as interstitials(vacancies) in the crystal\cite{OW,sqsoft,Sengupta,honey}. 
However, for hardcore bosons with nearest-neighbor repulsion, 
no supersolidity was found on the square\cite{sqhard, Hebert,Sengupta}, honeycomb\cite{honey}, kagome\cite{ka}, star lattice\cite{star} 
and Shastry-Sutherland\cite{ss} lattices, due to the instability of such a phase\cite{Sengupta,sqhard} on these 
lattices, which leads to phase separation into a pure solid and a superfluid for all values of the interaction strength.
The situation changes in the presence of geometric frustration\cite{Triangle1, Triangle2,Triangle3}.
Supersolidity was found for hardcore bosons on the triangular lattice, 
where the extensive degeneracy of the classical ground states at half filling is
lifted by quantum fluctuations and the ground state attains DLRO and ODLRO simultaneously, thus forming a
supersolid. 
Moreover, it was suggested that frustration induced by next-nearest neighbor interactions \cite{Kao} can also lead to 
supersolidity.  These findings show a different route to supersolidity, which is based on an order-by-disorder 
mechanism, by which a quantum system avoids classical frustration. 
However, all these investigations considered simple homogeneous lattices.
It is natural to investigate the behavior of bosons also on inhomogeneous lattices. 
\begin{figure}[b]
\begin{center}
\includegraphics[width=3.3in]{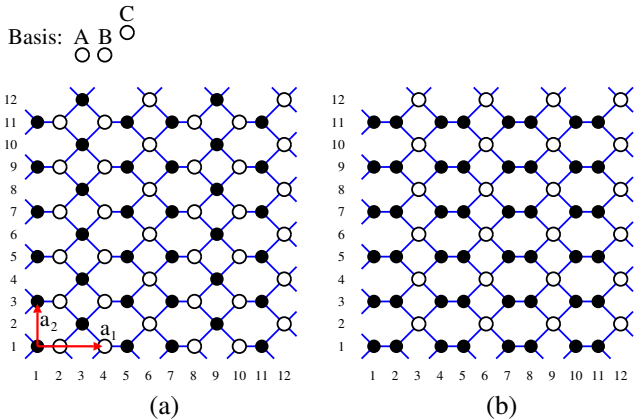}\\
\end{center}
\caption{(Color online). The dual of the bowtie lattice consisting of $n_1\times n_2=4 \times 6$ bases 
with periodic boundary conditions. The basis is shown in the top, where A, B and C label the three sublattices. The 
A and B sublattices are related by symmetry, and 
$\mathbf{a_1}$ and $\mathbf{a_2}$ are primitive vectors with length $3$ and $2$ respectively.
The numbers arranged in horizontal (vertical)  direction are the $x (y)$ coordinates
of the lattice sites.  The linear size of the system is $L_x=L_y=12$, with $N=L_x\times L_y/2=72$ sites. 
The coordination number of sites connected by horizontal bonds is 3, and that of sites 
with an $x$ coordinate satisfying $\mod(x,3)=0$ is 4.  (a) and (b) show the $\rho=1/2, 2/3$ SCDW phases respectively, 
where black circles represent bosons and white ones stand for vacant sites.}
\label{bow}
\end{figure}

Let us consider hardcore bosons with nearest-neighbor repulsion on the dual of the bowtie lattice, 
which is anisotropic in two directions and consists of two types of sites with different coordination number,  see Fig.~\ref{bow}.
The corresponding Bravais lattice is a rectangular lattice. It has a basis of 3 sites A, B, and C with the primitive vectors 
$\mathbf{a_1,a_2}$ as shown in Fig.~\ref{bow}. 
The Hamiltonian is
\begin{equation}
H=-t \sum_{\langle i,j \rangle} \left( a^\dagger_i a_j + a^\dagger_j a_i \right) 
  +V \sum_{\langle i,j \rangle} n_i n_j 
  -\mu \sum_i n_i,
\label{eqham}
\end{equation}
where $a^\dagger_i$ ($a_i$) creates (annihilates) a boson at site $i$,
$t$ is the nearest-neighbor hopping amplitude, 
$V$ the nearest-neighbor repulsion,  $\mu$ the chemical potential, and 
$n_i=0$ or $1$.

The model can be mapped onto the spin-1/2 XXZ model on the same lattice with $s_z(i)=n_i-1/2$
in the usual way,
\begin{equation}
H_s=-J \sum_{\langle i,j \rangle} (S_i^{+}S_j^{-}+S_i^{-}S_j^{+}) +J_z \sum_{\langle i,j \rangle}S_i^z S_j^z-\sum_i h(i) S_i^z,
\end{equation}
where $J=2t$ is the in-plane exchange, $J_z=V$ is the exchange in the $z$ direction, and $h(i)$ is 
a staggered external magnetic field given by  
$h(i)=\mu-2V$ for sites with coordination number 4 and $\mu-3V/2$, otherwise.
This marks the special character of the model. 
The solid state of the bosons is equivalent to magnetic order in the $z$ direction.


In the classical limit ($t=0$), at zero temperature, 
there are only two phases if the nearest-neighbor repulsion (or the exchange in $z$ direction) $V$ is absent. The 
lattice is empty when $\mu<0$, full if $\mu>0$.  

With the repulsion $V$ present, there exists four phases. For $\mu/V<0$, the lattice is empty.
For $\mu/V > 4$, the lattice is full. In the region $0<\mu/V<2$, the lattice is half filled. 
The model shows a solid ordering (antiferromagnetic order in $z$ direction) which breaks the translational 
symmetry of the original lattice, as shown in Fig. \ref{bow}(a).  
Two bases half filled with three bosons form a new basis of a solid at a wavevector $(\pi/3,0)$, 
showing the character of a striped solid or a SCDW\cite{roger}.  
We refer to this order as striped order I. 
In the region $2<\mu/V<4$ the lattice is 2/3 filled, showing a DLRO again. 
Two of three sites in a basis are filled and form the basis of  a phase with a 
wavevector $(2\pi/3,0)$, as shown in Fig. \ref{bow}(b). However, this ordering does not break
the translational symmetries of the original Bravais lattice, nor the symmetry of A and B sublattices. (The spontaneously breaking of 
the symmetry between two sublattices has been reported in \onlinecite{honey}).
Thus, we would not say the bosons form a solid. 
In this paper, we restrict a 'solid state' of lattice bosons as a diagonal long-range ordered state 
breaking either the translational symmetry of the original lattice or the symmetry of different sublattices. 
The order is again 
of a striped or a SCDW type, to which we refer as striped order II.
With quantum hopping present, SCDW phases were found in models with next-nearest neighbor repulsion\cite{roger} or 
plaquette interactions\cite{sand}.  For hardcore bosons on the dual of the bowtie lattice,
without the next-nearest neighbor repulsion or plaquette interactions, we shall show that striped phases also emerge.

\begin{figure}[t]
\includegraphics[width=2.5in]{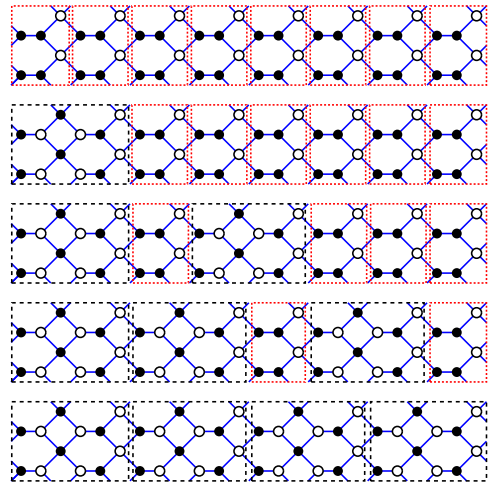}
\caption{ (Color online). The possible boson (black dot) distributions on the $24\times4$ ($n_b=8$) dual of the bowtie lattice at $t=0, 
\mu=2V$  with density 12/24, 13/24, 14/24, 15/24, 16/24 (from bottom to top, $n$ from 4 to 0) respectively. 
}
\label{distribution}
\end{figure}

At $\mu/V=2$, a special degeneracy of the ground states appears, as illustrated in Fig.~\ref{distribution}. 
Since breaking one of the two stripe ordered patterns along the $y$-direction costs energy,
no interface can be formed along the horizontal direction at zero temperature. 
The lattice can be divided into blocks along the $x$-direction without creating any such interfaces. 
The length of a half filled block in the $x$-direction 
is $6$, that of the 2/3 filled one is $3$. 
For a system with linear size $L=3 n_b$ in the $x$-direction ($n_b$ is the number of bases along the $x$-direction, which is 
restricted to be even in this work), 
let $n$ and $n_b-2 n$ be the number of blocks with particle density $\rho=1/2$ or $2/3$ respectively,
the energy density (per site) of the system is  $E=(-3 n \mu+ (n_b-2 n)(-2 \mu + V))/(3n_b)=-V$,
which is independent on $n$.   
Thus, there are $n_b/2+1$ possible boson densities: $\rho=(2n_b- n)/(3n_b), n=0,1,\cdots,n_b/2$.
For each filling, there is a total of $N_d(\rho)=n_b (n_b-n-1)!/(n_b-2n)!/n!$ degenerate states. 
In other words, there are $n_b/2+1$ phases coexisting. In the thermodynamic limit,
the system can have any density between 1/2 and 2/3.
Although the zero temperature entropy per site is still 0 in the thermodynamic limit,  
the long-range order is broken in these phases, except for the half filling and the 2/3 filling phases.  
In the language of the spin-1/2 XXZ model,
the long-range magnetic order is destroyed by the competition between the staggered field and the exchange in the $z$ direction, 
instead of the geometrical frustration.
This is different from the frustrated ground state on the triangular lattice at half filling.

Considering these interesting properties of the model introduced by the inhomogeneity and the anisotropy of the dual of the bowtie lattice, 
it is desirable to explore quantum properties of the model.
To this end, we performed extensive simulations 
by using the stochastic series expansion (SSE) quantum Monte Carlo method 
with directed loop updates\cite{sseloop} for the hardcore bosons on the dual of the bowtie lattice. 

{\it quantum phase diagram}

\begin{figure}[t]
\includegraphics[width=2.5in]{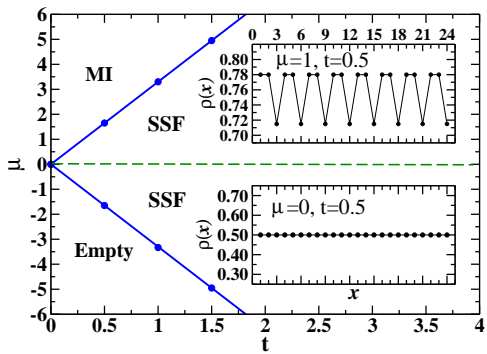}
\caption{(Color online). Phase diagram at the limit $V=0$.  The upper inset shows density wave: particle density per column  as a function 
of the $x$ coordinate of the column, at $\mu=1, t=0.5$. The density wave is absent at $\mu=0$, as shown in the lower inset. }
\label{phdV0}
\end{figure}

We start with the simple limit $V=0$. The phase diagram is shown in Fig. \ref{phdV0}.
We find that the system is in a phase with striped order II and nonzero superfluid stiffness between 
a Mott insulating phase (MI) and an empty phase, except for $\mu=0$ where the model has an exact particle-hole symmetry.

\begin{figure}[b]
\begin{center}
\includegraphics[scale=0.32]{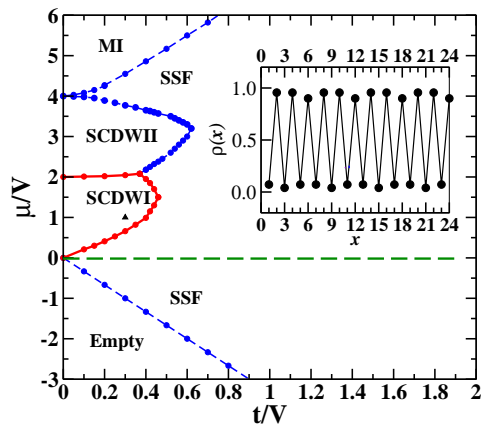}
\caption{(Color online). Phase diagram with nearest-neighbor repulsion present. Blue (dashed) lines indicate continuous phase 
transitions, and the red (bold solid) line represents first order transitions.
The inset shows the density wave in the SCDW I phase at $\mu/V=1, t/V=0.3$, indicated by the filled uptriangle. }
\label{phd}
\end{center}
\end{figure}

The phase diagram for $V \neq 0$ is much richer, see Fig.~\ref{phd}.
We find SCDW I with $\rho=1/2$ and  SCDW II with $\rho=2/3$ 
when the hopping is weak, as expected. 
The phase transition between the two SCDW phases is first order. 
A stable SSF emerges outside the two SCDW phases, except for $\mu=0$.
We will now proceed to discuss the phase diagrams in more detail. 

To characterize density wave order, one usually measures the static structure factor $S(\mathbf{Q})$ with $\mathbf{Q}$ the wavevector 
\begin{equation}
S(\mathbf{Q})/N=\frac{1}{N^2} \langle (\sum_{k} n_k e^{i \mathbf{Q}\cdot \mathbf{r}_k})^2\rangle,
\end{equation}
where $k$ labels sites, and $N$ is the total number of sites.
However, in our case, it is more convenient to distinguish the two striped phases 
by introducing the following quasi-structure factors, which measure the density differences between different sublattices,
or, in the language of the XXZ model, the various magnetic orders in the $z$ direction: 
\begin{equation}
S_{\alpha}/N=\frac{1}{N^2} \langle (\sum _{i} e^{(\alpha)}_{i} n_{i})^2 \rangle.
\end{equation}
For $\alpha=1$, $e^{(1)}_i=(-1)^{x(i)}$, where 
$x(i)$ is the $x$ coordinate of site $i$.
$S_1/N$ measures the square of the density difference between the two sublattices consisting of sites with even or odd $x$ coordinate,
showing a density wave at wave vector $(\pi/3,0)$ if not zero.
For $\alpha=2$, $e^{(2)}_i=1$, if $\mod (x(i),3)=1$; $e^{(2)}_i=-1$, if $\mod(x(i),3)=2$; $e^{(2)}_i=0$, otherwise.
$S_2/N$ is the square of the density difference between the A and B sublattices. 
We also define the third quasi-structure factor 
\begin{equation}
S_3=\frac{1}{N}\langle \sum_i e^{(3)}_i n_i \rangle,
\end{equation}
with $e^{(3)}_i=0$, if $\mod(x(i),3)=1$; $e^{(3)}_i=1$, if $\mod(x(i),3)=2$; $e^{(3)}_i=-1$, otherwise. 
$S_3$ measures the difference of the boson densities on the B and C sublattices. 
In the present model, we find that $S_2$ is always zero, which means that the symmetry between the A and B sublattices is always kept, thus
$S_3$ is also the density difference of the bosons on the A and C sublattices.

The superfluid stiffness is measured in terms of winding number fluctuations \cite{sf},
\begin{equation}
\rho_{s} ^{\alpha}=\frac{ <W_{\alpha}^{2}>}{ \beta t}\, ,
\end{equation}
where $\alpha$ labels the  $x$ or $y$-direction, and $\beta$ is the inverse temperature. 
Typically the superfluid density $\rho_s$ is the average of the two stiffnesses.
Considering the anisotropy between the $x$ and $y$-directions, 
we shall distinguish the superfluid stiffness in the two directions.

We measure the particle density, the quasi-structure factors and the superfluid stiffnesses as
functions of the chemical potential $\mu$ and the hopping amplitude $t$ (in units of the nearest-neighbor repulsion $V$,
if $V \neq 0$).
In the simulations, we set $L_x=L_y=L$, which is restricted to multiples of 6. The total number of sites is $N=L^2/2$.
The inverse temperature was chosen as $\beta=2 L/t$ to make sure that the simulations access the ground state
properties. 

{\it quantum phase diagram for $V=0$.}

We first consider the non-repulsive limit $V=0$.  The phase diagram is shown in Fig.~\ref{phdV0}, 
which is invariant under the interchange of particles with holes $\rho \rightarrow 1-\rho$ 
and the change of sign $\mu \rightarrow -\mu$.
The phase boundaries can be well predicted in the single particle picture. The system is empty 
when  $\mu < -10t/3$, and full when $\mu > 10t/3$, where $10/3$ is the average coordination number (twice the ratio of the total numbers of
bonds and sites).  Our simulations confirm this prediction.

\begin{figure}[tpb]
\includegraphics[scale=0.35]{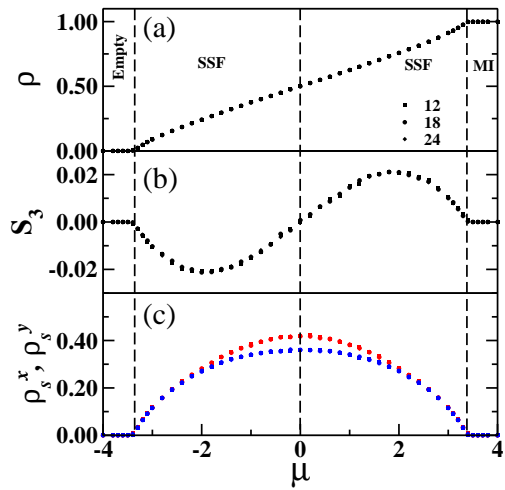}
\caption{(Color online). The particle density $\rho$,  quasi-structure factor $S_3$, and superfluid stiffnesses $\rho^x_s$ 
(red (light) symbols) and $\rho^y_s$ (blue (deep) symbols) 
as functions of $\mu$ at  $V=0, t=1$.} 
\label{v0mu}
\end{figure}

Fig.~\ref{v0mu}(a) shows the boson density $\rho$ as a function of the chemical potential $\mu$ with hopping $t=1$.
We see that the density varies continuously from 0 to 1 as $\mu$ changes from -10/3 to 10/3.
In this region, $\rho_s^x$ and $\rho_s^y$ become nonzero with $S_3 >0$ for $\mu >0$, or $S_3<0$ for $\mu <0$, as shown in 
Fig.~\ref{v0mu}(b) and (c). 
This means the system displays both DLRO and ODLRO. 
To further confirm this, we sample the average particle density at the $x$-th column  
$\rho(x)=\langle \sum_{i=1}^{L_y/2} n_i(x) /(L_y/2)\rangle$, $i$ denotes  sites in the $x$-th column. 
A density wave at wavevector $(2\pi/3, 0)$ is seen in the upper inset of Fig. \ref{phdV0} for $\mu=1, t=0.5$. The density wave 
for $\mu=-1, t=0.5$ follows by applying the transformation $\rho \rightarrow 1-\rho$. 
For $\mu=0$, no such wave appears, as shown in the lower inset.
Thus the system indeed has a striped order of type II.  

This result can be understood in the following way: 
At $\mu=10t/3$, holes can appear in the system. Each hole costs a potential energy $\mu$, 
and gains kinetic energy $-10 t/3$ by hopping freely. 
The wave functions of the holes spread out over the entire lattice, but the probability of finding a hole in the C sublattice is 4/3 times 
of that in the B (or A) sublattice due to the coordination number difference. 
The striped density wave thus starts to develop.
As $\mu$ decreases, the price of creating a hole becomes cheaper and the density of holes becomes larger, thus
the density wave is getting stronger. As the hole density further increases, the hardcore nature of the bosons prevents 
further increase of the density difference between the C and A (or B) sublattices. 
The difference approaches a maximum, 
and then starts to decrease.
Finally, at $\mu=0$, the lattice is half filled and the difference disappears.
However, as mentioned above, this striped order II does not break the translational symmetry of the 
original lattice, nor the symmetry between the A and B sublattices. 
Thus, we would call this phase bearing both ODLRO and DLRO a striped superfluid (SSF), instead of a supersolid phase (SS).  
This SSF phase is a result of the interplay of the chemical potential and the lattice inhomogeneity and anisotropy.

This picture can also be understood in the language of the XXZ model. 
At $\mu=10t/3$, the $xy$ components of spins start to align, forming a ferromagnetic phase due to the competition between the 
in-plane coupling and the external magnetic field $h=\mu$.  As the field $\mu$ decreases, 
the ferromagnetic order in the $xy$ plane gets stronger, and meanwhile the magnetization in the $z$ direction decreases. 
The inhomogeneity of the lattice structure makes the local magnetization of $z$ component in C sublattice weaker than that in 
the A and B sublattices. This results in a staggered long-range ferromagnetic order in the $z$ direction. 
When the external field $\mu$ reaches 0, the magnetization in the $z$ direction disappears completely.

An interesting phenomenon in this model is that the superfluid stiffness along the $x$-direction (perpendicular to the stripes) 
becomes larger than 
that along the $y$-direction (parallel to the stripes) near half filling. 
In contrast, a larger superfluid stiffness along the stripes than transverse to the stripes was reported in various striped 
supersolid states on the square and triangular lattices\cite{sqsoft, Hebert, Kao, ST, roger}, where the lattice itself is
isotropic.  
Clearly, the anisotropy of the superfluid stiffness reflects the inequivalence of the $x$ and $y$-directions of the dual of 
the bowtie lattice, but it is not clear a priori that which of them should be larger. 

{\it quantum phase diagram for $V \neq 0$.}

\begin{figure}[tb]
\begin{center}
\includegraphics[scale=0.35]{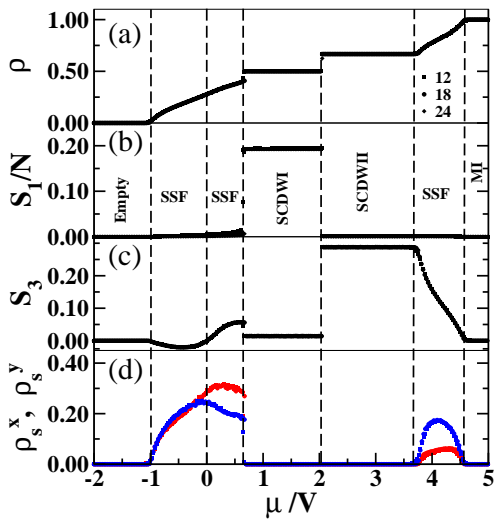}\\
\caption{(Color online). The particle density $\rho$, 
quasi-structure factors $S_1/N, S_3$, and superfluid stiffnesses $\rho_s^x$ (red (light) symbols) and $\rho_s^y$ (blue (deep) symbols) 
as functions of $\mu/V$, at the cut $t/V=0.3$. }
\label{ob1}
\end{center}
\end{figure}

With the nearest-neighbor repulsion $V$ present, we have a different phase diagram, without the symmetry when interchanging particles
with holes. 
In the single-boson picture, the system becomes empty when  $\mu \le -10t/3$. 
Our simulation results for $V>0$ are in agreement with this picture.  
Considering a single freely hopping hole on the lattice, one can show that the system sits in the 
$\rho=1$ state when $\mu \ge 10 (t + V)/3$, which is the Mott insulating state.
This is true for $t/V \gg 1$, where the single hole added is almost free with its wavefunction spreading over the whole lattice.  
Our simulations show that the MI boundary is straight with the expected slope in this region. 
However, the boundary is curved when the system is approaching the classical transition point $\mu/V=4, t=0$.  
For small $t/V$, holes are created on the C sublattice, which costs the same chemical potential $\mu$ but gains a potential energy $4V$, 
more than for those created on the A or B sublattices. These holes are forced to hop along $y$-direction. 
The kinetic energy gain of a single hole  by the second order hopping processes is $-4t^2/V$. 
Otherwise, the kinetic energy gain $t$ cannot compensate the cost of the potential energy.
This explains the stronger superfluid stiffness along the $y$-direction, see Fig.~\ref{ob1}(d), and results in the curved 
boundary: $\mu = 4V +4t^2/V$.

We now show results obtained by scanning the chemical potential at constant $t/V$. 
In Figs.~\ref{ob1}(a)-(d), we plot the particle density $\rho$, the quasi-structure factors $S_1/N, S_3$ and the superfluid 
stiffnesses $\rho_s^{x},\rho_s^{y}$ as functions of the chemical potential $\mu/V$ for $t/V=0.3$ respectively. 
$S_2/N$ is always zero, which means that the boson densities in A and B sublattices are always equal.  

Between $\mu/V=2$ and $3.7$, the $\rho=2/3$ SCDW II phase is found with superfluid density $\rho_s=0$, which means that the bosons
are localized. 
The quasi-structure factor $S_3 \approx 0.28$, which is less than the value $1/3$ for the exact static SCDW II phase due to the 
presence of hopping $t$.

At $\mu/V=3.7$, the density starts to grow continuously as $\mu$ increases, indicating a second order phase transition.
$S_3$, $\rho_s^{x}$ and $\rho_s^{y}$ are all finite in the region $ 3.7< \mu/V < 4.6$.
The model displays both DLRO and ODLRO.
Again, since the striped order II does not break the translational symmetry of the original lattice,
the model turns out to be in a SSF phase, which can be understood in terms of doping bosons in the SCDW II state.
The doped bosons are mobile along the $y$-direction in the C sublattice and generate stronger superfluid stiffness along the $y$-direction 
than along the $x$-direction. 
The doped bosons can not form a domain wall which breaks the striped order II. Thus the SSF phase is stable.
Equivalently, the phase can be thought as doping holes in the Mott insulating state, which hop along the $y$-direction to avoid 
potential energy cost and form a stronger superfluid stiffness in $y$-direction, as described in previous text.

A SCDW I phase is clearly observed in the region $0.68 <\mu/V <2$, with $\rho=1/2$, 
$S_1/N=0.193$, $S_3=0.014$, and $\rho_s^{x},\rho_s^{y}$ converging to 0. 
Here, the striped order I shown in Fig. \ref{bow} (a) is slightly adapted by an additional order II.
We show the density per row $\rho(x)$ in the inset of Fig.~\ref{phd}. The ordering 
at wavevector $(\pi/3,0)$, which breaks the translational symmetry of the original lattice, 
is clearly seen.  Thus the bosons here form a true symmetry-broken solid state.

\begin{figure}
\includegraphics[width=3.0in]{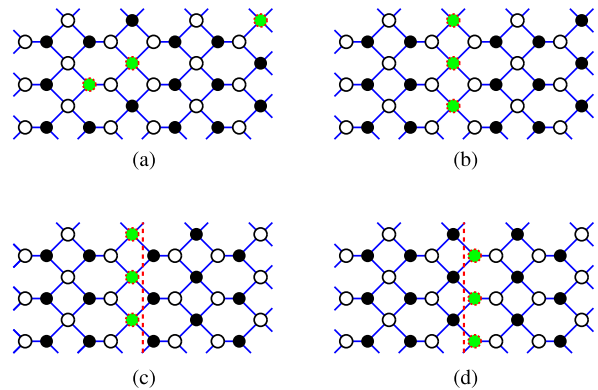}
\caption{(Color online). The $\rho=1/2$ solid doped with holes. 
(a) Holes (green circles with dashed boundary) added in the solid. (b) Lining the holes costs no additional energy. (c) A domain wall (dashed red line) is introduced 
at no cost by shifting the right half of the lattice.  (d) The holes can hop freely across the domain wall and gain kinetic energy.}
\label{stable}
\end{figure}
Doping this solid (SCDW I) with holes leads to a phase separation, instead of supersolidity. 
A density jump caused by a first order phase transition is observed at $\mu/V=0.68$, leaving SCDW I for a $\rho < 1/2$ superfluid phase. 
The mechanism is similar to what was found for the square lattice and the triangular lattice\cite{Sengupta,Triangle1}. 
Adding $L/2$ holes onto the SCDW I solid 
decreases the density infinitesimally in the thermodynamic limit. Each
hole costs a chemical potential $\mu$ and gains a kinetic energy which is quadratic in $t$ 
by the second-order hopping processes (see Fig.~\ref{stable}(a)).
Placing these holes along a vertical line costs no additional energy, as shown in Fig.~\ref{stable}(b).
We can shift one half of the lattice to the left (or right) by one unit of $x$ coordinate, introducing
a domain wall which breaks the SCDW I order without cost of energy, as shown in Fig.~\ref{stable}(c). 
By hopping freely across the domain wall, each additional hole
gains a kinetic energy $-t$, which lowers the energy of the domain wall state compared to the bulk supersolid (see Fig.~\ref{stable}(d)).
Thus, the supersolid phase with SCDW I solid order is unstable. 
However, the striped order II due to the inhomogeneity and the anisotropy of the dual of the bowtie lattice 
is still present when the boson density $\rho$ is lower than 1/2. We see that $S_3$, which shows the density difference 
between B (or A) and C sublattices, is not zero (Fig. \ref{ob1}(c)). 
Since the superfluid density is also nonzero in this region,
the system is actually in a SSF phase,  which finally ends when the lattice becomes empty at $\mu/V=10(t/V)/3$, 
with the line $\mu=0$ as an exception.
Thus the phase separation at $\rho=1/2$ is between a SCDW I phase and a SSF phase.

Moreover, we notice that $\rho_s^x$ is much stronger than $\rho_s^y$ when $\rho$ is close to 0.5, where $S_3\approx 0.05$, which is much 
larger than its maximum value at $V=0$. This means that bosons are partly localized due to the presence of a repulsion $V$. 
The localized boson density on the A (or B) sublattice
is larger than that on the C sublattice. Holes generated in the A (or B) sublattice contribute to the superfluid stiffness along the $x$-direction, 
which is transverse to the stripes. This anisotropy is again very different from what found in various striped supersolid states near half 
filling\cite{sqsoft, Hebert,Kao, ST,roger}.
\begin{figure}
\centering
\includegraphics[width=2.5in]{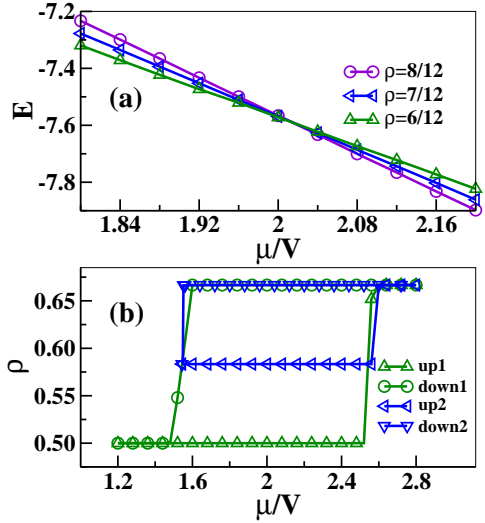}
\caption{ (Color online). Energy level crossing (a) and multi-loop hysteresis (b) for hardcore bosons on a $12\times 12$ lattice at
$t/V=0.2$.  Three phases with different densities coexist near $\mu/V=2$.  
}
\label{1stopt12}
\end{figure}

The phase transition between SCDW I and SCDW II is rather special.
As discussed above, at the transition point $\mu/V=2$, there are actually $L/6+1$ coexisting phases in the classical limit.
With hopping $t$ nonzero, but small, we can still see that these states are degenerate  near  $\mu/V=2$, as shown in Fig.~\ref{1stopt12} (a).
Away from, but close to, $\mu/V=2$, one of the states becomes the ground state, the others are metastable.
This situation is illustrated by the interesting multi-loop hysteresis curves for a $12 \times 12$ system 
at $t/V=0.2$, see Fig.~\ref{1stopt12} (b).
During the simulations,  we store the last configuration of a finished simulation as the initial configuration of the next simulation
with a new value of the chemical potential\cite{roger, mss}.
Starting from $\mu/V$ much less than 2 to ensure that the system stays in the ground state, i.e. $\rho=1/2$ SCDW I, we increase $\mu/V$ and sample the
particle density $\rho$.
The system  does not jump to the real ground state ($\rho=2/3$) immediately when $\mu/V$ passes the 
transition point $\mu/V=2$. This curve is labeled as 'up1' in Fig.~\ref{1stopt12}(b).
Then, we start with $\mu/V \gg 2$ to ensure that the system stays in the $2/3$ filling SCDW II state, then decrease $\mu/V$. 
The system may jump to the $\rho=1/2$ state which is the ground state, closing a hysteresis loop (the curve 'down1'),
or jump to a metastable phase with 7/12 filling by chance (the curve 'down2'),  when $\mu/V$ is small enough. 
We can use the latter configuration as the initial configuration, and increase $\mu/V$ again. It is seen that the system stays in
the metastable state until $\mu/V$ reaches a value much larger than 2, as shown by the curve 'up2'. Curves 'down2' and 'up2' form another
hysteresis loop.

This interesting phenomenon makes the simulations near $\mu/V=2$ very difficult. The data near this point, shown in Fig.~\ref{ob1}, 
are obtained by initializing configurations in the 'right' way.  
The transition points between the SCDW I and SCDW II phases can be found from the energy level crossings.

\begin{figure}
\includegraphics[scale=0.35]{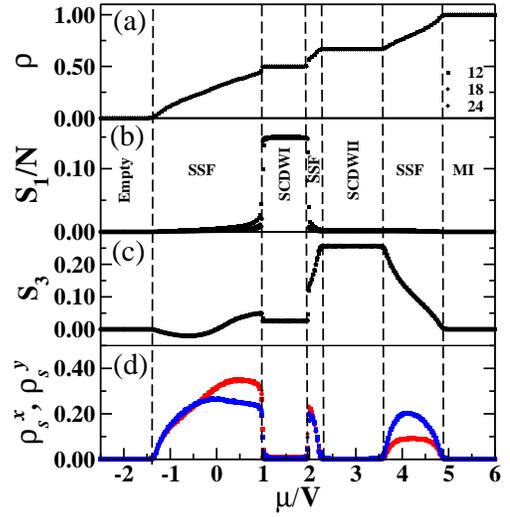}
\caption{(Color online). The particle density $\rho$, 
quasi-structure factors $S_1/N, S_3$, and superfluid stiffnesses $\rho_s^x$ (red (light) symbols) and $\rho_s^y$ (blue (deep) symbols) 
as functions of $\mu/V$, at the cut $t/V=0.4$. }
\label{obt1v2.5}
\end{figure}

Increasing hopping $t/V$ larger than $0.37$, we find that the SSF phase emerges in the region between the two SCDW phases.
To demonstrate this, we take $t/V=0.4$, and scan the chemical potential $\mu$. 
The particle density $\rho$, quasi-structure factors $S_1/N, S_3$, and superfluid stiffness $\rho_s^x$ and $\rho_s^y$ are plotted in 
Fig. \ref{obt1v2.5}(a)-(d) respectively.
Between the SCDW I and SCDW II phases, we see that striped order II and superfluidity coexist.
It is also clear that the transition from SCDW I to the SSF phase is of 
first order, and the transition from SCDW II to the SSF phase is continuous. 
In this region, the superfluid stiffnesses in the two directions become different with $\rho_s^x > \rho_s^y$ near half filling.
This is also the case for the SSF state with $\rho < 1/2 $ near half filling. For the SSF state with $2/3<\rho<1$,
we see the same anisotropic behavior as that at the cut $t/V=0.3$.

\begin{figure}[tb]
\begin{center}
\includegraphics[scale=0.35]{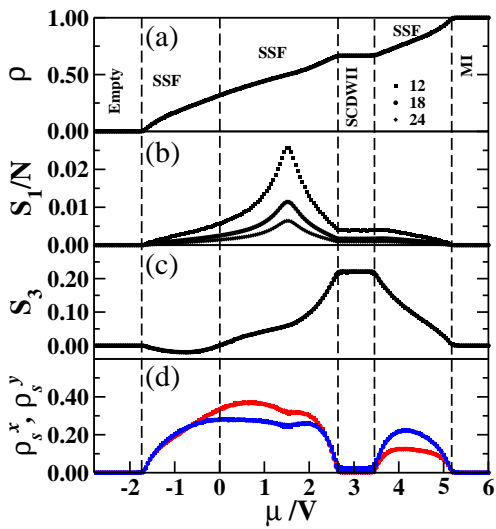}\\
\caption{(Color online). The particle density $\rho$, 
quasi-structure factors $S_1/N, S_3$, and superfluid stiffnesses $\rho_s^x$ (red (light) symbols) and $\rho_s^y$ (blue (deep) symbols) 
as functions of $\mu/V$, at the cut $t/V=0.5$. }
\label{ob2}
\end{center}
\end{figure}

The general picture of the phase diagram along $\mu/V$ for a cut at constant $t/V$ changes as $t/V$ becomes even larger. 
At $t/V=0.5$ shown in Fig.~\ref{ob2}, the 2/3 filling SCDW II phase is the only density wave phase.
Although the finite size data of $S_1/N$ are nonzero in a large region, a finite-size scaling analysis indicates that they 
finally converge to zero as system size turns to infinity.  The strong hopping destroys the SCDW I order. 
The SSF state with both DLRO and ODLRO is found outside the SCDW II phase
for $\mu/V < 2.65$ and $\mu/V >3.5$.
Close to, but outside the SCDW II phase, it is seen that the type II striped order persists. 
Meanwhile, the superfluid stiffnesses start to increase.
Hence, the SSF state appears.
We still see a large anisotropy of superfluid stiffness in the SSF states: $\rho_s^x > \rho_s^y$ near half filling, and 
$\rho_s^y > \rho_s^x$ for $2/3 <\rho <1$. 

{\it conclusion}

We have investigated the ground state behavior of hardcore bosons on the dual of the bowtie lattice. 
A special state (SSF) with a striped order II (see Fig. \ref{bow}) and an ODLRO is found even in the absence of repulsion. 
The emergence of such a state is due to the inhomogeneity and the anisotropy of the lattice structure. 
However, neither the translational symmetry nor 
the symmetry between the A and B sublattices of the original lattice is broken.
Such SSF states should exist on anisotropic and inhomogeneous lattice with sites having different coordination numbers.
Including a nearest-neighbor repulsion causes a much richer phase diagram.
Two SCDW phases with different striped order are found.  A SSF phase exists outside the two SCDW phases, between the Mott 
insulating phase and the empty phase.
Comparing with the SSF state with $V=0$,  we see that the striped order II is enhanced greatly by the repulsion in the SSF state.
The phase transition between the two SCDW phases is first order. 
The transition between the SCDW I and SSF state is also first order, while the transition between SCDW II and SSF state is continuous.
We also report anisotropies of superfluid stiffness in the SSF states with and without repulsion.
In the SSF state with repulsion, the superfluid stiffness shows different anisotropies in the region near half filling and 
above 2/3-filling.
Hardcore bosons on inhomogeneous lattice present rich and interesting properties.
Further theoretical  or experimental studies are worthy to explore the nature of the SSF state in more detail.  

\acknowledgments
We thank A. W. Sandvik, Y.-J. Kao, H. W. J. Bl\"ote, S.-J. Yang for valuable discussions.
The work  is supported by the NSFC under Grant No. 10675021, and by the HSCC (High Performance
Scientific Computing Center) of the Beijing Normal University.

\end{document}